\documentclass[10pt,conference,fleqn]{IEEEtran}
\IEEEoverridecommandlockouts

\usepackage{cite}
\usepackage{amsmath,amssymb,amsfonts}
\usepackage{graphicx}
\usepackage{caption,subcaption}
\usepackage[table,xcdraw]{xcolor}
\usepackage{textcomp}
\usepackage{algorithmic}
\usepackage[linesnumbered,ruled,vlined]{algorithm2e}
\usepackage{adjustbox}
\usepackage{multirow}
\usepackage{multicol}
\usepackage{booktabs}
\usepackage{bigstrut}
\usepackage{verbatim}
\usepackage{array}
\usepackage{rotating}
\usepackage{xcolor}
\usepackage[a4paper, total={184mm,239mm}]{geometry}
\def\BibTeX{{\rm B\kern-.05em{\sc i\kern-.025em b}\kern-.08em
    T\kern-.1667em\lower.7ex\hbox{E}\kern-.125emX}}
\begin{document}
\title{A Joint Optimization of Buffer and Splitter Insertion for Phase-Skipping Adiabatic Quantum-Flux-Parametron Circuits}
\author{\IEEEauthorblockN{Robert S. Aviles, Peter A. Beerel}
\IEEEauthorblockA{\textit{Department of Electrical and Computer Engineering, University of Southern California}, Los Angeles, USA \\\{rsaviles, pabeerel\}@usc.edu}}

\maketitle

\begin{abstract}
Adiabatic Quantum-Flux-Parametron (AQFP) logic is a promising emerging device technology with six orders of magnitude lower power than CMOS. However, AQFP is challenged by the fact that every gate must be clocked, where proper data transfer requires connected gates to have shifted but overlapping clocks.  As a result, buffers need to be used to balance re-convergent logic paths, a problem that is exacerbated by every multi-node fanout needing a tree of clocked splitters. Recent AQFP circuit design techniques have offered an opportunity to reduce buffer costs by supporting a notion of phase-skipping but the EDA support for these advanced circuits is limited.  This paper proposes the first algorithm to optimize buffer and splitter insertion for phase-skipping AQFP circuits and achieves over 31\% savings over existing buffer reduction schemes and up to 74\% savings in buffers and splitter costs over the SOTA non-phase skipping circuits. 
\end{abstract}

\section{Introduction}
Computing performance has traditionally enjoyed rapid improvements in energy and area efficiency thanks to the shrinking of transistor sizes following Moore's law \cite{moore} in \textit{complementary metal-oxide-semiconductor} (CMOS) technology.  However, after decades of consistent improvements in semiconductor fabrication, transistor sizes are reaching a practical minimum \cite{end_moore}. This looming end to Moore's Law, coupled with increasing power densities and a practical limit on clock frequencies for CMOS devices, has driven the demand for emerging device technologies to provide energy-efficient high-performance alternatives. Due to exceptional energy-efficiency and high frequencies, superconductive \textit{single-flux quantum} (SFQ) \cite{isvlsi2} logic devices have become a promising replacement for CMOS in a range of applications from space to exasale supercomputing.  

\begin{figure}[t]
\includegraphics[width=0.8\columnwidth]{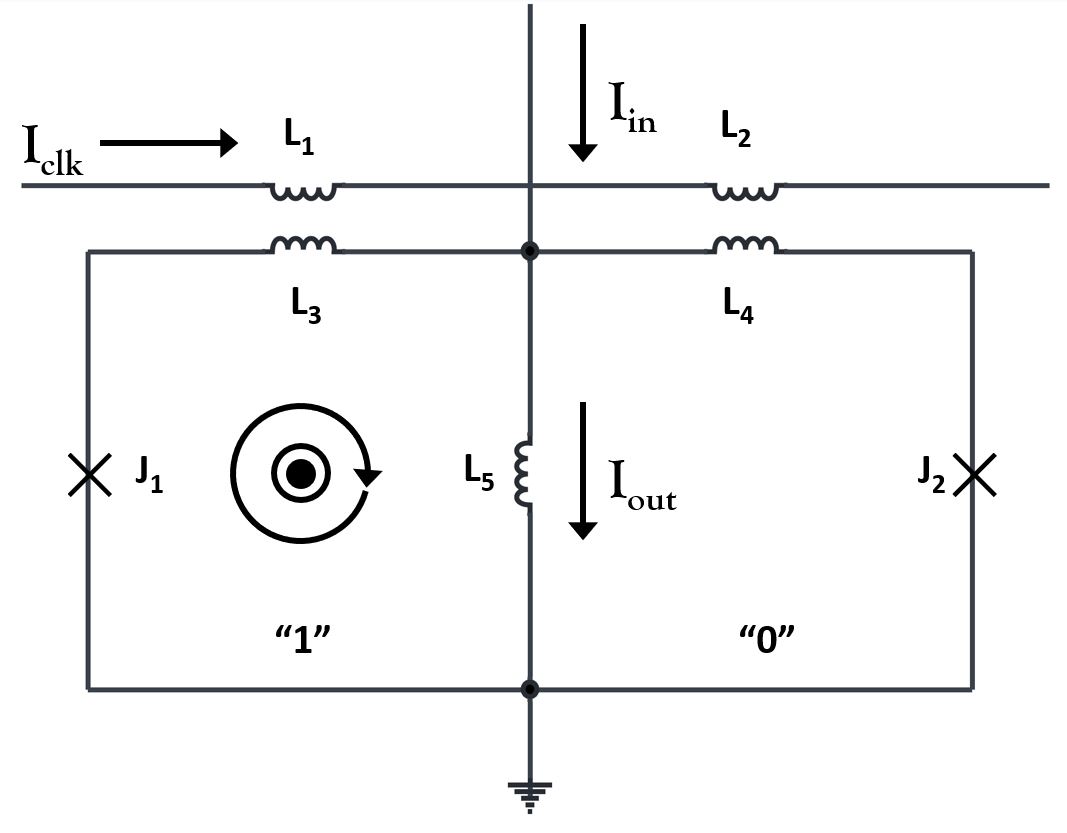}
\centering
\caption{Circuit Construction of AQFP Buffer.  Current directions shown for when flux is in the left loop, corresponding to a logical "1".}
\label{buf}
\end{figure}

Cooled to 4.2K, Rapid Single Flux Quantum (RSFQ) devices operate at hundreds of GigaHertz with switching energy on the order of $10^{-19}$ J.  However, RSFQ devices require constant DC bias currents which can lead to high static power dissipation.  To address this, numerous superconductive designs have been proposed to minimize static power dissipation including efficient rapid single flux quantum (ERSFQ)\cite{ersfq}, efficient SFQ (eSFQ) \cite{esfq}, reciprocal quantum logic (RQL) \cite{rql},  low-voltage RSFQ \cite{lv-rsfq}, LR-biased RSFQ \cite{lr-rsfq}, and \textit{Adiabatic Quantum Flux Parametron} (AQFP) \cite{AQFP}. AQFP is amongst the most promising of these logic families. AQFP avoids the inherent static power dissipation issues associated with DC-biased RSFQ logic types, consuming zero static power \cite{zero_static}. 
It does this using AC-currents in superconducting wires to excite logic gates that act as both power and clock (see Fig.~\ref{buf}). 
Even accounting for the energy overhead to cool to 4.2K, AQFP achieves two orders of magnitude advantage in EDP compared with state of the art semiconductor technologies \cite{cooling_overhead}.  Moreover, the recent breakthroughs in higher temperature superconducting operating at 25K \cite{HighTC} dramatically increase the range of potential applications for this technology.


One limitation of AQFP is that every gate must be clocked, where proper data transfer requires connected gates to have shifted but overlapping clocks. This constraint is traditionally satisfied using a 4-phase AC clocking signal in which neighboring gates are connected to successive phases. In large circuits with unbalanced paths, this constraint requires the insertion of buffers to balance re-convergent logic paths. Exacerbating this problem is that wires need clocked splitter cells to support a fanout of greater than one which can increase the need for path-balancing buffers. In extreme cases path-balancing buffers can occupy over 90\% of the circuit \cite{buffer_costs}. 
\begin{figure*}[t]
\includegraphics[width=2\columnwidth]{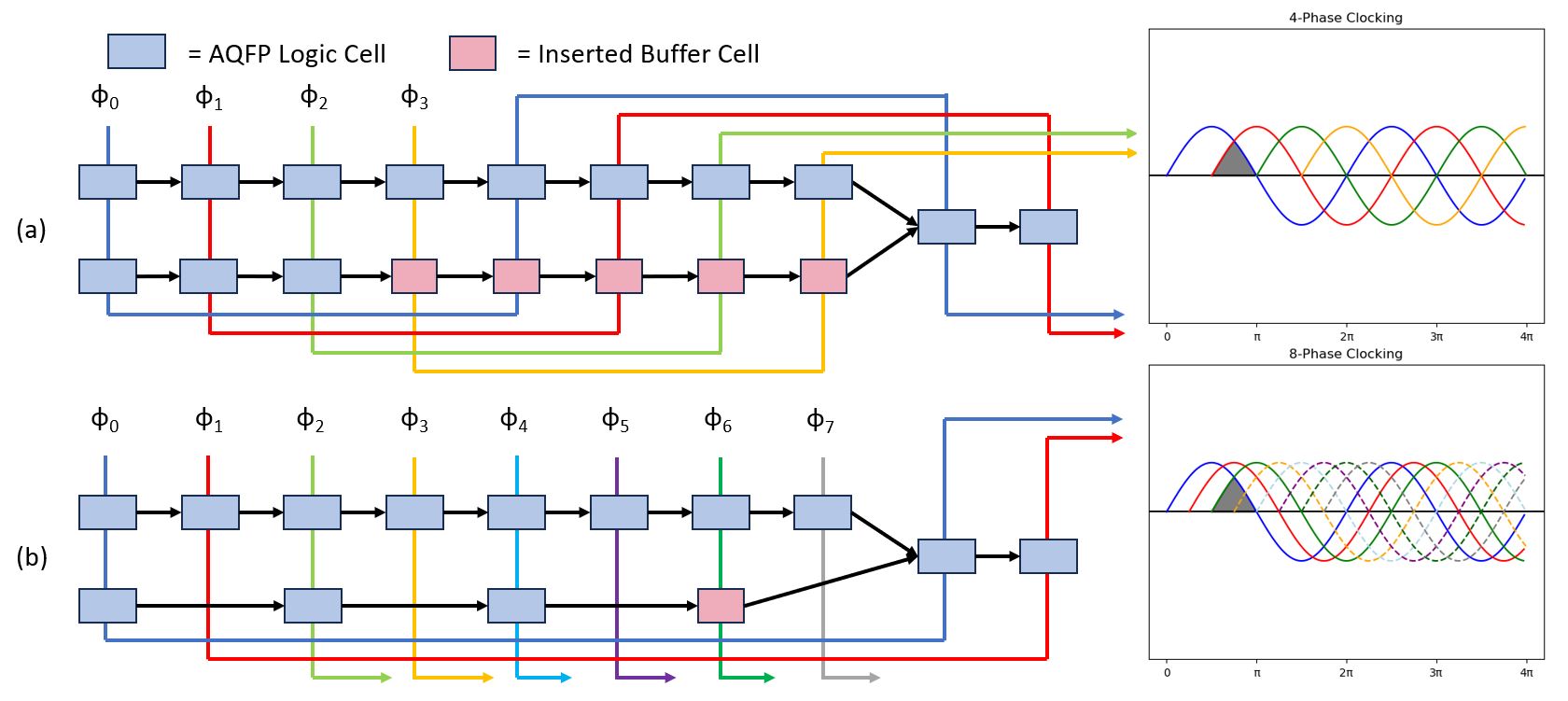}
\centering
\caption{Example clock routing and waveforms for (a) 4-Phase and (b) n-Phase schemes 
(n=8). Shaded area illustrates that for 8-phase clocking $\phi_0$ has equivalent overlap with $\phi_2$ as with $\phi_1$ in 4-Phase, allowing a phase skip and the associated reduction in inserted buffer cells.  In (b) Phase-Skipping aware phase assignment of logic cells reduces inserted buffer costs.}
\label{layout}
\end{figure*}
To mitigate these effects and improve circuit integration, considerable effort in electronic design automation (EDA) has been made towards minimization of inserted buffer and splitter costs \cite{dp_OG,bs-2,heuristicASP,DAC22,SOTA}. 
Despite these algorithmic improvements the buffer and splitter (B/S) overhead still accounts for the majority of AQFP circuit area. Alternative works have shown that removing path balancing constraints by repeating input patterns can lead to dramatic area improvements at the cost of significant drop in throughput \cite{phaseMatch}.
More recently, new AQFP circuit design techniques \cite{Nphase,delayln1} have 
offered an opportunity to reduce buffer costs by supporting a notion of phase-skipping. 
These new clocking schemes enable overlap between a range of non-adjacent clock phases, relaxing the strict path balancing constraints and promising improved integration density.

The solution space for B/S optimization for phase-skipping circuits is much larger than for circuits that require adjacent clock phase assignments. To address this challenge this paper makes the following contributions.

\begin{itemize}
    \item We propose the first buffer and splitter (B/S) insertion algorithm that can exploit phase-skipping clocking schemes achieving 31.6\% improvement in B/S costs over existing buffer reduction techniques.
    \item Our algorithm demonstrates that circuits that support skipping 4 phases can be optimized to achieve a 74.1\% reduction in buffer and splitter cost over circuits that do not support phase-skipping, which leads to 40.7\% reduction in total JJ count.
    \item Each iteration of our algorithm includes an optimal phase-skipping enhanced level assignment algorithm as an integer linear program (ILP) which we can approximate as an LP.  Moreover, we couple this algorithm with a polynomial-time optimal splitter tree construction presented in \cite{dp_OG,SOTA}
    extended to support phase-skipping circuits.
    \end{itemize}

The remainder of this paper is organized as follows.  Section \ref{background} provides background information on AQFP Superconducting Logic and the underlying circuit improvements that have motivated these new EDA algorithms.   Sections \ref{algooverview} and \ref{algodetails} provide an overview and details of our proposed algorithm, respectively.  Section \ref{experiments} presents our experimental results quantifying our improvements over the SOTA. 

\section{Background}\label{background}

\subsection{AQFP Superconducting Logic} 

AQFP buffers encode a logic state by the location of a single flux quantum within one of two superconducting loops. The choice of loop determines the direction of the output current flow, with the positive direction being considered a logical '1' and the negative direction a '0', as illustrated in Fig.~\ref{buf}. The location of this internal flux is set by the direction (logic state) of the input current flow.  Basic logic functionality is implemented with Majority-3 logic, where the output of a 3-input branch (merge) cell is set by the sum of 3 input currents \cite{scl}.  Since  NOT logic can be achieved by inverting buffer outputs using appropriately designed transformers, a universal set of functions can be realized from the Majority-3 gate \cite{maj3}.  More precisely, AQFP circuits are typically implemented using 4 basic building blocks, the AQFP buffer, NOT, a constant cell, and branch cells to achieve merge and splitter functionality \cite{scl}.   


AQFP devices generate output current only when receiving a positive AC excitation current that is coupled to its superconducting loops via inductors, labeled as $I_{clk}$ in Fig. \ref{buf}.  Moreover, the input current must be present at least a setup time prior to the rise of the AC excitation current \cite{Nphase}.  
In this manner the excitation current behaves similar to a clock for each cell, where inputs are processed on the positive clock edge and the output is driven only while the clock signal is active. Accordingly, proper operation requires that clock phases for connected gates are offset to satisfy setup constraints but have sufficient active overlap to ensure proper data transfer \cite{overlap}.  Traditionally, this has been realized using 3-phase and 4-phase clocking, where each clock has the same frequency but with shifted phase. Typically, each gate in a given level is driven by the same clock phase and the clock lines are snaked throughout the design as shown in Fig. \ref{layout}.  Accordingly, phase-assignment is derived from the gates assigned logic level and the two terms are often used interchangeably.

\subsection{Phase-Skipping Clock Schemes}

Interestingly, due to the high inherent switching speed of AQFP, the circuits can easily meet the setup constraint with achievable 4-phase clock frequencies. Thus motivated, n-phase clocking \cite{Nphase} and delay-line clocking \cite{delayln1} have been proposed both yielding smaller phase shifts where each active clock phase now has overlap with multiple clock phases.  As a result, both of these methods allow connections between gates driven by non-adjacent phases, referred to as phase skipping.   

N-phase clocking is illustrated in Fig. \ref{layout}(b) where the shaded region indicates significant overlap between $\phi_0$ and $\phi_2$ allowing direct connection without buffer insertion at $\phi_1$. With n-phase clocking a driving gate may directly connect to gates with $\lfloor\frac{N}{4}\rfloor$ subsequent phases \cite{Nphase}. For example, calculations show that up to 20 phases may be implemented at a 5 GHz frequency \cite{Nphase}, enabling $\lfloor\frac{20}{4}\rfloor - 1 = 4$ phases to be skipped at each connection. 
In addition to phase skipping, n-phase clocking leads to lower latency. For example, 20-phases reduces the latency of a 4-phase design by $5x$ by enabling data to transition through up to 20 logic gates instead of four in the same clock period.   


N-phase clocking can be implemented using externally generated phases or generated internally through delay-line clocking \cite{delayln1}.  Using delay-line clocking a phase skipping of four phases has been demonstrated \cite{delayln2}. 
The EDP of an AQFP gate with delay-line clocking was shown to be 2.8 x $10^{-32}$J$\cdot$ s, compared to 1.4 x $10^{-31}$J$\cdot$ s for 4-phase clocking, and 6.8 x $10^{-29}$J$\cdot$ s for RSFQ \cite{delayln2}.

Both implementations of n-phase clocking enable reduction of existing netlists through removal of buffers that are within the permitted phase-skipping range \cite{Nphase}. However, such buffer chain reduction only operates on existing netlists where splitter tree structures and phase assignments are not optimized to exploit phase-skipping.  Our work constructs optimized netlists for phase-skipping schemes to maximally reduce the total buffer and splitter cost of a circuit.  

\section{Proposed Iterative Algorithm Overview}\label{algooverview}

\begin{figure}[t]
\includegraphics[width=0.8\columnwidth]{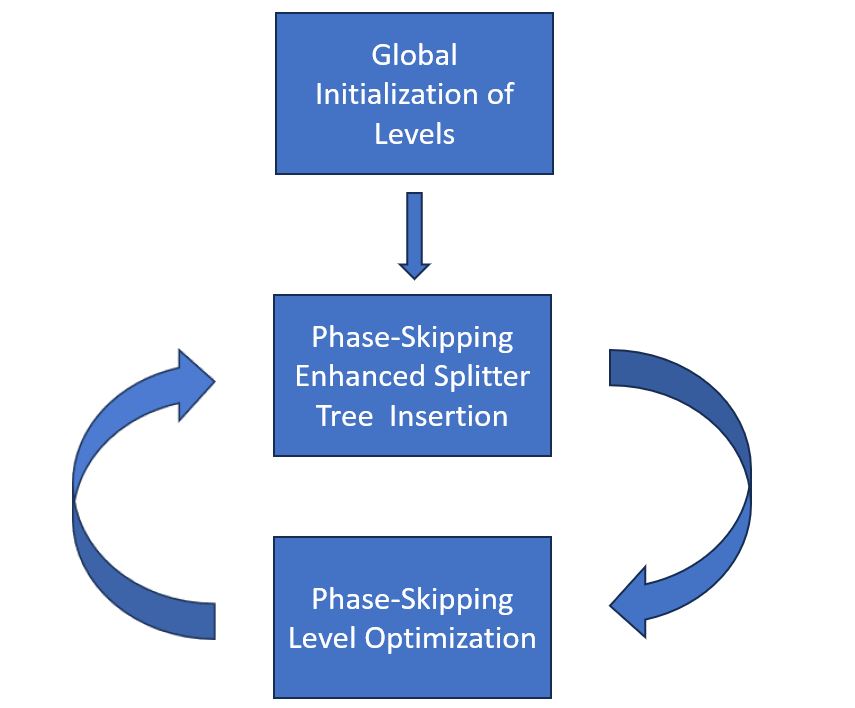}
\centering
\caption{Example Flow of Proposed Iterative Algorithm}
\label{flow}
\end{figure}

In AQFP, since all splitters are clocked elements, the main challenge of minimizing buffer and splitter insertion is that the choice of splitter tree topology affects the degree of imbalance between reconvergent paths and constrains the optimal level assignment solution.  Similarly, a fixed level assignment for gates may constrain the available splitter tree configurations that can be inserted. Accordingly, it is difficult to optimize the level assignments and the splitter tree configurations as both rely on knowledge of the other.  To address this challenge we propose an iterative solution (Fig. \ref{flow}), switching between globally optimizing level assignments and locally optimizing splitter trees, with each step enhanced to exploit phase-skipping connections.  
In the first step we heuristically initialize level assignments using a
polynomial complexity linear program (LP) that captures the allowable phase skipping and includes a constraint from \cite{SOTA} that estimates the size of the needed splitter trees. 

After each level assignment we replace existing splitter trees
with trees that are locally optimized for the new assigned fanout levels. 
Optimal splitter tree construction for a single fanout tree has been well researched with a polynomial complexity dynamic programming implementation presented in \cite{dp_OG,SOTA}.  This splitter tree algorithm takes fanout level assignments as inputs and returns the splitter tree construction that minimizes, in order of priority:  the maximum increase in level assignment, the total increase in level assignment, and the total buffer/splitter costs.  We modify this dynamic program to include phase-skipping aware optimization.  Additionally, we extend this algorithm to utilize a notion of allowable slack, removing penalty for beneficial level re-assignments during splitter tree construction.  

Our iterative flow optimizes level assignment with the existing splitter tree topology, and then adapts that topology to the level assignments, where each step of our algorithm improves upon the global solution quality.  This is because the \textit{i}th level assignment step is constrained by the resulting topology of the splitter tree construction in iteration $i-1$.  Similarly, the splitter tree construction is constrained by the level assignments of the previous step. As a result, we can guarantee that after level assignment, a valid splitter tree configuration exists for the given level assignments.  Accordingly, the splitter tree insertion for the level assignments will construct a new splitter tree with lower or the same local cost ensuring that each local splitter tree insertion either improves our global solution or returns a solution of the same quality. Similarly, for a given splitter tree the level assignment step will improve the solution quality or return the same solution.  Our algorithm then not only improves upon solution quality with each iteration, but also returns feasible netlist solutions at each step. We allow our algorithm to run until no improvement is found between iterations.  Across our benchmarks we 
never exceeded 15 iterations for a circuit.

\section{Iterative Algorithm Details}\label{algodetails}

\subsection{Phase-Skipping Enhanced Level Optimization}\label{initlvl}

With inserted splitters treated as nodes, we formulate the clock phase assignment problem of both splitters and gates as an optimization problem minimizing the total number of buffers inserted. We model the circuit as a directed acyclic graph (DAG) $(V,E)$ where each gate is an internal node in the graph and circuit inputs (outputs) are input (output) nodes. A directed edge ($E_{ij} \in E$) is created between each pair of nodes in $V$ that is connected in the circuit. The number of buffers to be inserted between connected nodes $i$ and $j$ is represented as $C_{ij}$. Each node is assigned a level $L_i$ which directly corresponds to its clock phase assignment. Accordingly, any level difference between connected nodes that exceeds the given circuit's phase skipping constraints requires a buffer insertion.  We create constraint (\ref{eq:cij_cost}) to capture the required buffer cost for an allowable overlap of adjacent phases $P$, such that $P=2$ for circuits with a phase skip of 1.  Following standard design constraints, all primary inputs (I) are assigned level $1$ and all primary outputs (O) are forced to have equal levels.  This leads to the introduction of the following optimization problem: 

\begin{equation} \label{eq:newcost}
    \text{Minimize:} \sum_{E_{ij}=1}{C_{ij}}
\end{equation}
\begin{equation*}
    \text{subject to:}
\end{equation*}
\begin{equation}\label{eq:cij_cost}
 1\leq L_j-L_i \leq {(C_{ij}+1)*P} \hspace*{0.3in}  \forall (i,j) \in E_{i,j},
\end{equation}
\begin{equation}
 L_i = L_{outputs}  \hspace*{0.3in} \forall i \in O, 
\end{equation}
\begin{equation}
    L_i = 1    \hspace*{0.3in} \forall i \in I,
\end{equation}
\begin{equation}
\label{eq:Integral}
    L_i, C_{ij} \in \mathbb{N} \hspace*{0.3in}  \forall (i,j) \in E_{i,j}
\end{equation}

When integer constraints are enforced (\ref{eq:Integral}) we can directly account for all buffers being inserted, optimally assigning the levels for the given splitter tree and phase-skipping constraints. 

After a solution is returned, the inserted buffer cost plus the number of presently inserted splitters is the total solution cost.   

\subsubsection{Linear Relaxation}\label{approx}

Considering that the optimal level assignments for a given splitter tree are typically not the final level assignments and the optimal levels will change as the splitter tree configuration adapts, it is worthwhile to relax the integer constraints of the ILP, achieving scalability necessary for successive iterations at a minor sacrifice in solution quality.  To handle non-integer values, all returned \textit{$L_i$} values are rounded up to the nearest whole number. Since all values are rounded in the same direction, connected nodes that were subject to the relationship $L_j-L_i\geq{1}$ before rounding, will still satisfy the constraint after rounding.  Some non-optimality is introduced as our objective function may now minimize the sum of fractional buffers ($C_{ij}$) when $\lceil C_{ij}\rceil$ buffers must be inserted. Importantly, a given iteration may now have a minimal increase in buffer count due to this approximation, in which case the algorithm stops and the previous iterations solution is returned.  Our experimental results demonstrate that this relaxation yields strong performance. 

\subsubsection{Initial Heuristic Assignment}

For only our iteration start point, before splitter insertion, we heuristically estimate splitter tree levels during an initial level assignment. To do this, we add a constraint taken from \cite{SOTA} that provides lower bounds on the difference between levels of connected gates, as follows. 
\begin{equation}
\label{eq:borrowed}
    \sum_{j \in S}{(L_j - L_i - 1) \geq f(|S|)} \hspace*{0.3in} \forall i \in v, \forall S \in RS_i 
\end{equation}
This constraint was inspired by the observation that for any subset \textit{S} of fanout nodes, the {\em all path sum} of the associated splitter tree must be greater than or equal to that of the minimal all path sum $f(|S|)$ of a complete balanced \textit{X}-way tree with $|S|$ leaves $B_{|S|}$ (where \textit{X} is the maximum fanout of a splitter). Because $B_{|S|}$ is a tree, $f(|S|)$ is equal to the sum of the depths of each leaf of $B_{|S|}$. 

Put more simply, the levels of any combination $S$ of fanout nodes must leave room for the minimum depth splitter tree that connects the source node to those $|S|$ fanouts.   Accordingly, the sum of level differences between a source node and any subset of its fanouts must be greater than the smallest all path sum of a splitter tree that can satify the given fanout requirements.

However, for a node with $t$ fanouts, enumerating all possible subsets requires $2^t$ combinations which becomes prohibitively large.  In \cite{SOTA} they fully enumerate trees with up to 30 fanouts which can lead to up to $2^{30}$ (over 1 billion) constraints being added to their ILP for a single splitter tree. Since we are only using constraint (\ref{eq:borrowed}) for our initial level assignment, we can use a much lower number of constraints $K$ in our formulation. In particular, we set our limit to fully enumerate 14 leaf nodes which corresponds to a maximum of $K = 16,384$ subsets. For nodes with fanouts larger than 14, we randomly generate $K$ subsets. The set of subsets is labelled as $RS_i$ in Eq. \ref{eq:borrowed}. Compared to \cite{SOTA}, we can employ this more relaxed set of constraints as well as use a simpler LP instead of an ILP, because the subsequent steps of our algorithm improve upon the initial level assignment resulting from this formulation.  
Additionally, during initialization we modify the objective function (Eq. \ref{eq:initial}) such that the local costs $C_{ij}$ are divided by the number of fanouts of node $i$ as a heuristic estimate of potential buffer sharing in the splitter tree.
\begin{equation} \label{eq:initial}
    \text{Minimize:} \sum_{E_{ij}=1}{\frac{C_{ij}}{|Fanouts_i|}}
\end{equation}
Both this modified objective function and constraint (\ref{eq:borrowed}) are only used during our initial level assignment.

\subsection{Phase-Skipping Enhanced Splitter Tree}\label{splitteralgo}

\subsubsection{Algorithmic Principles}

We leverage and extend the dynamic-program based optimal splitter tree insertion algorithm presented in \cite{dp_OG}. In particular, this algorithm is originally designed for circuits without phase-skipping with the objective to minimize the maximum and total extra delay, where extra delay is the required increase in level assignment for a given node to accommodate splitter tree insertion.

In general, this algorithm works by exploring a 4-dimensional array in which an array point describes how to decompose the optimal solution of a problem into two sub-problems. The key observation behind the algorithm is that ordering the leaf nodes by increasing level assignments enables an optimal solution be built up from combining minimum cost subtrees that {\em span a range} of leaves. 
\cite{dp_OG}. This fixed leaf node ordering shrinks the solution space and enables an optimal dynamic programming solution \cite{dp_OG}. 


To enable phase-skipping we have to modify the calculation of buffer costs and expand the set of subtrees that should be considered when using them to construct a larger subtree.  More specifically, the buffer cost to connect to a splitter or leaf is the level difference divided by the maximum phase span of connections (maximum phase-skipping + 1) rounded down.  Phase-skipping allows construction between subtrees at different levels without extra costs. In the original formulation, a subtree could only consider combining solutions at the next level (depth + 1).  However, with phase-skipping, solutions can be combined from any combination of solutions from levels within the phase-skipping range (depth + 1 + maximum phase-skip).  This modification requires an extra FOR loop when exploring minimum cost solutions to sub-problems,  increasing the algorithmic complexity by a factor of phase-skip which is fixed for a given design.  

The original proof of optimality \cite{dp_OG}only applies to a solution space for which all level re-assignment is minimized. Accordingly, the original formulation ensures an increase in level assignment (extra delay) occurs only when the assigned levels cannot accommodate the insertion of a feasible splitter tree.  However, \cite{dp_OG} does not consider the fact that level adjustment can often improve both local and global solution quality. Therefore in addition to phase-skipping modifications we extend the tree insertion algorithm to include a notion of allowable slack in the extra delay. While including allowable slack improves overall solution quality in the circuit, optimality is not guaranteed.  This is because level re-assignment can disrupt the ascending order of the leaf nodes, implying that an optimal utilization of slack might require a different ordering of leaf nodes.  However, since this slack utilization can only improve the solution we accept this potential for non-optimality, and leave algorithms that guarantee optimal solutions for future work.

The allowable slack of a fanout node is defined as the difference between a gate's currently assigned level and the level of the last buffer connected to the gate's output in the current solution.  This slack allows buffers to be retimed across the gate without counting as extra delay in our cost function.  Similar to classical retiming methods \cite{leiserson1991retiming}, the gate can take the level assignment of the last buffer with no additional cost. Because the affects of buffer retiming do not propagate to other nodes, the splitter tree is allowed to re-adjust the gates level assignment within the allowable slack if it improves the local solution quality. 
However, it is critical that slack is only allowed for 2-input gates (realized as Majority-3 gates with a constant input), since, as in classical CMOS, retiming a buffer across a gate causes duplicate buffers to be created on each of the gates inputs. Therefore, retiming a 2-input gate leads to global change of $+1$ buffers. However, this extra delay will only occur if it improves the local splitter tree on one of the gates fanin's, in effect absorbing at least 1 retimed buffers within the local splitter tree, guaranteeing better or equivalent global buffer cost.  This feature of allowable slack permits our splitter tree to consider a range of local level re-assignments to improve both local and global solution quality.  


\subsubsection{Implementation Details}

\begin{figure}[t!]
\begin{subfigure}[t]{\columnwidth}
  \includegraphics[width=1.0\linewidth]
{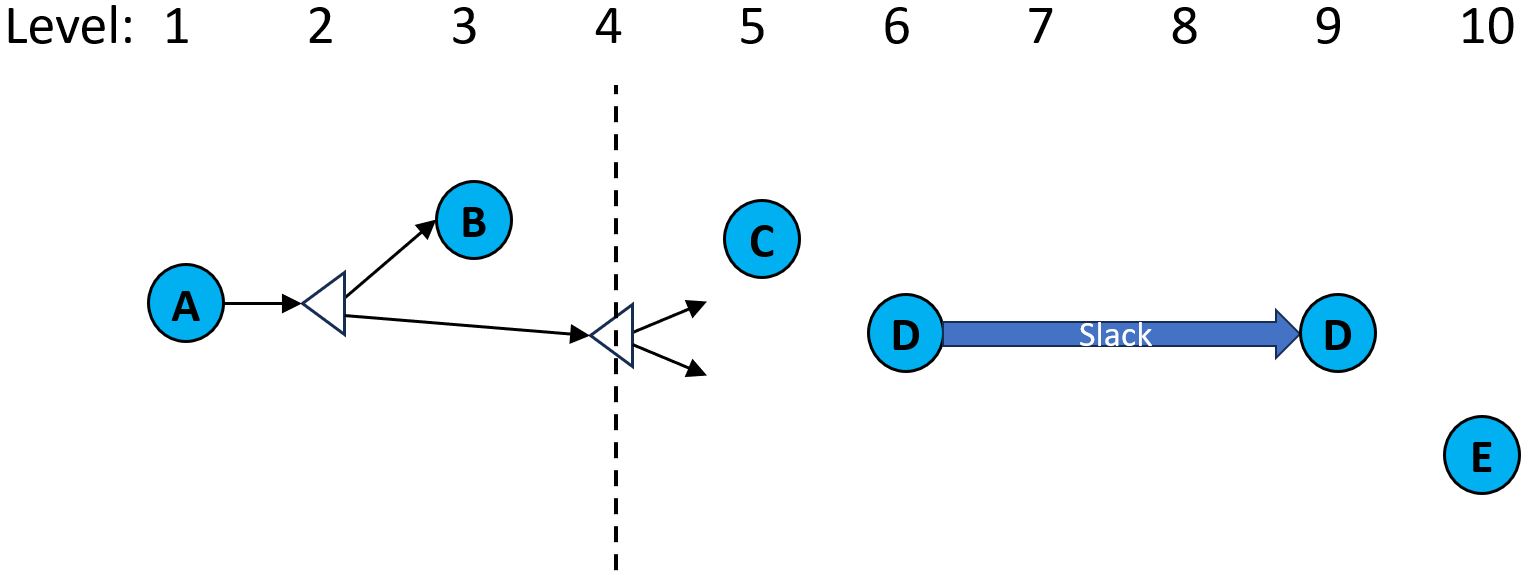}   \caption{Example construction at $pt[C][E][2][4]=\langle C,1,4\rangle$ at dashed line.  The array indices correspond to constructing the minimum cost subtree containing leaf nodes $[C]$ through $[E]$, with the base of the substructure having $[2]$ fanouts at level $[4]$.} 
  \label{fig:ptstart}
\end{subfigure}\\
~
\begin{subfigure}[t]{\columnwidth}
  \includegraphics[width=1.0\linewidth]{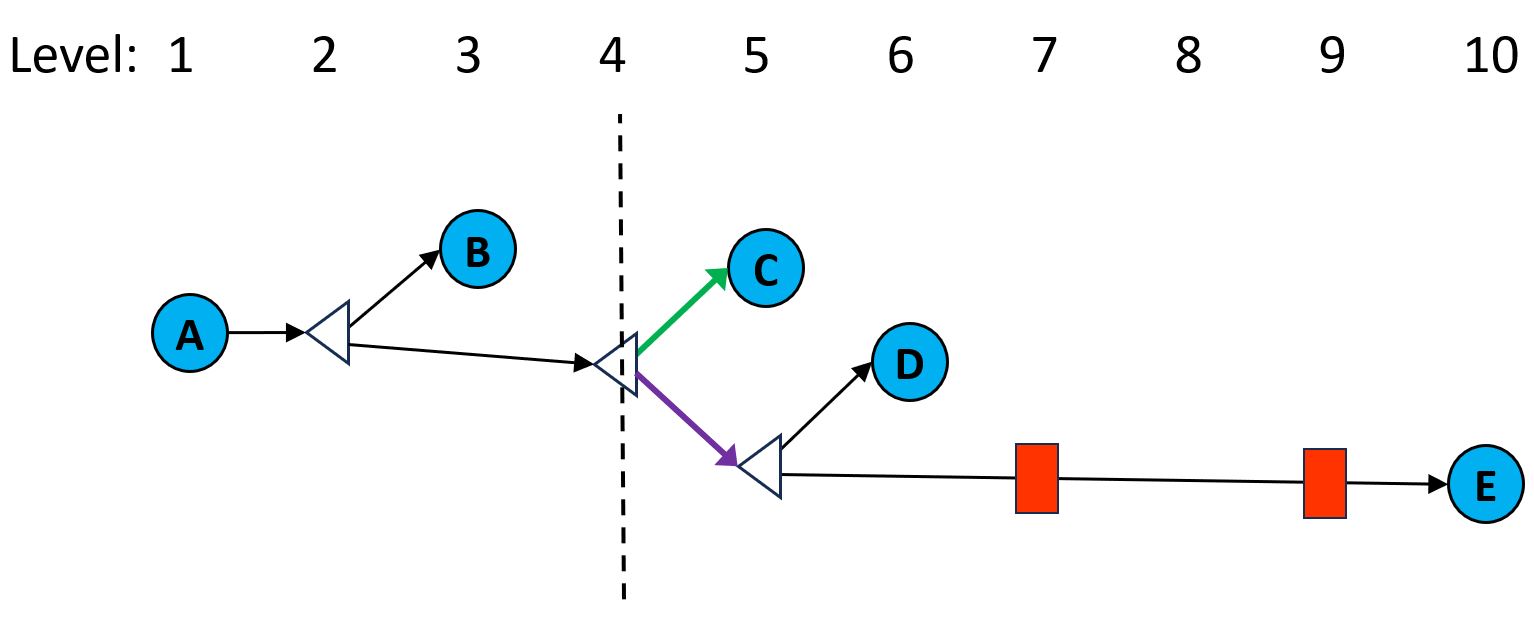}  
  \caption{Minimal tree without slack utilization, with inserted buffers shown in red. Subtree costs are $dp[C][C][1][4]=\langle0,0,0\rangle$ (green) and $dp[D][E][1][4]=\langle0,0,3\rangle$ (purple) for total cost $\langle0,0,3\rangle$.}
  \label{fig:ptworse}
\end{subfigure}\\

~
\begin{subfigure}[t]{\columnwidth}
  \includegraphics[width=1.0\linewidth]{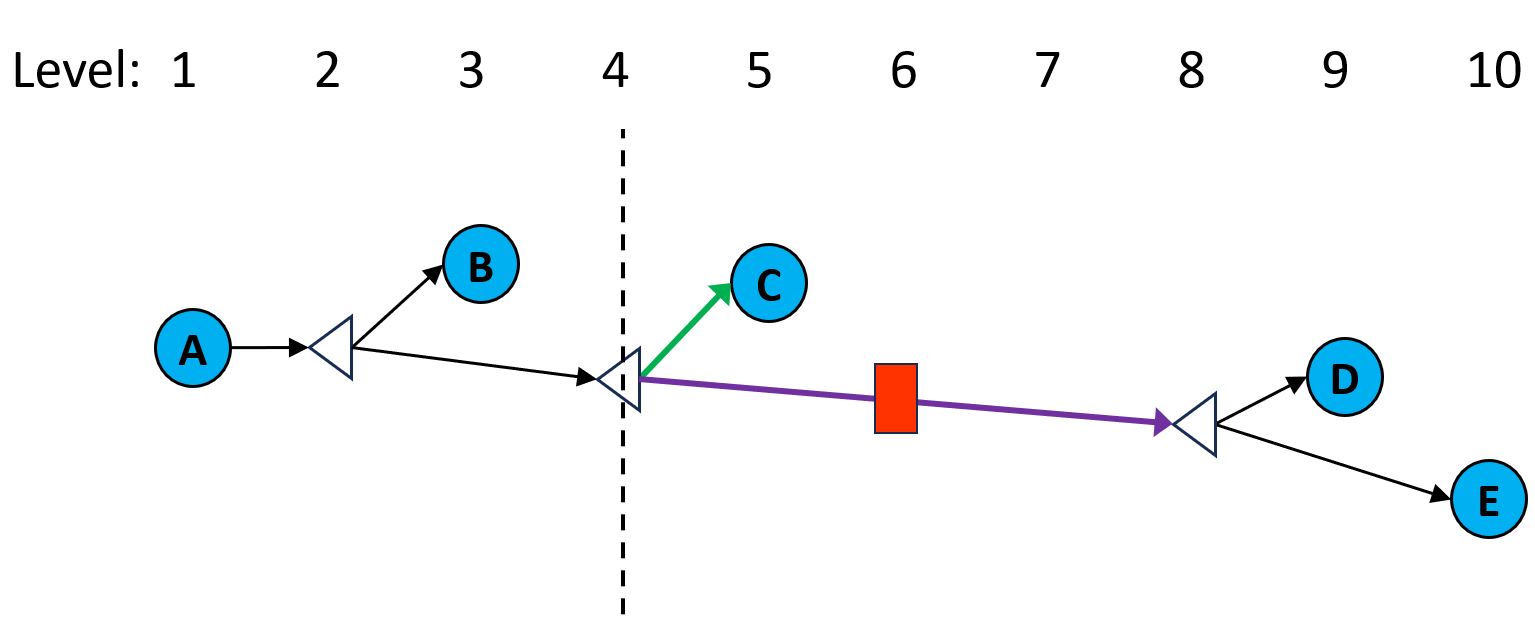}  
  \caption{Utilizing slack leads to increased buffer sharing reducing local cost. $dp[C][C][1][4]=\langle0,0,0\rangle$ (green) and $dp[D][E][1][4]=\langle0,0,2\rangle$ (purple) for total cost $\langle0,0,2\rangle$.}
  \label{fig:ptopt}
\end{subfigure}
\caption{Example of splitter tree subtree constructions for phase-skipping of 1 with a splitter fanout of 2.}
\label{fig:pt}
\end{figure}

Our resulting dynamic program is presented in Algorithm \ref{splitters}. We follow the formulation from \cite{SOTA}, where $dp_{l,r,b,d}$ records the cost for a tree rooted at level $d$ with $b$ branches that contains leftmost leaf $l$ to rightmost leaf $r$ as \textit{\{maximum extra delay, total extra delay, total buffer and splitter cost\}}, and the cost is minimized in descending order of priority where maximum extra delay is given highest precedence. By building up from subtrees containing a single leaf, the minimum cost subtree for each set of leafs can be constructed allowing the minimal cost tree satisfying fanout to all leaf nodes to be constructed. 

\begin{algorithm}
\SetKwInput{KwData}{Input}
\SetKwInput{KwResult}{Output}
\caption{PhaseSkippingTreeInsertion($F,X,ps$)\label{splitters}}
\KwData{$F$ the set of $n$ fanouts of the given source node $s$ ordered by ascending delay $(\delta)$ where $F_i.\delta = L(F_i)-L(source)-1$, $F_i.a$ is the allowable slack on a node, $X$ the maximum fanout for a splitter, and $ps$ the maximum phase span of connections for the given circuit }
\KwResult{$pt$ the pivot table that can re-construct the minimum cost splitter tree through backtracking}
  $D_{max} \gets F_n.\delta + \lceil log_X n \rceil$\;
  \tcc{Initialize direct connections}
  \For{$b \in [1,min\{n,X\}]$; $l \in [1,n-b+1]$}
  {
  $r \gets l+b-1$\;
  $\delta \gets F_l.\delta$\;
  \eIf{b=1}
  {
  \tcc{Calculate cost to connect to $l$ from all depths}
  \For{$d \in [0,D_{max}]$}
  {
  $\Delta \gets |d-\delta|$\;
  \eIf{$d>\delta$}
  {
  $\textbf{dp}_{l,l,1,d} \gets d>\delta+F_l.a$  ? $\{\Delta,\Delta,0\}: \{0,0,0\}$\;\label{slack}
  }{
  $\textbf{dp}_{l,l,1,d} \gets \{0,0,\lfloor\frac{\Delta}{ps}\rfloor\}$\;\label{cost}
  }
  }
  }
  {
   \tcc{Calculate cost to connect to $l$ through $r$ from splitter at max level}
  $\textbf{dp}_{l,r,b,d} \gets \textbf{dp}_{l,r-1,b-1,D_{max}}$+$\textbf{dp}_{r,r,1,D_{max}}$\;
  }
  }
  \tcc{Find minimum cost for all sub-problems}
  \For{$ln \in [1,n-1]$; $l \in [1,n-ln]$}
  {
  $r = l + ln$\;
  \For{d = $D_{max}-1$ to 0; $s \in [1,min(X,ln+1)]$}
  {
  \eIf{$b=1$}
  {
    \tcc{Find minimum cost connection to buffer or splitter}
  $\textbf{dp}_{l,r,b,d}\gets \displaystyle\min_{\substack{\forall k \in [1,min(X,ln+1)]\\\forall pr \in [1,min(ps,D_{max}-d)]}} \textbf{dp}_{l,r,k,d+pr}$+1\;\label{span}
  $\textbf{pt}_{l,r,b,d}\gets (-1,k,d+pr)$\;\label{pt1}
  }{
    \tcc{Construct set of branches from 2 solved sub-problems with minimum costs}
  \For{$k \in [1,b-1]$}
  {
  $\textbf{dp}_{l,r,b,d}\gets \displaystyle\min_{\forall p \in [l,r-b+k]}$ $\textbf{dp}_{l,p,k,d}$ + $\textbf{dp}_{p+1,r,b-k,d}$\;
  $\textbf{pt}_{l,r,b,d} \gets (p,k,d)$\;\label{pt2}
  }
  }
  }
  }
  \Return {$\textbf{pt}$}\; 
\end{algorithm}


The cost for all possible subtrees is explored in \textbf{dp} and our pivot table (\textbf{pt}) is recursively traversed to construct the splitter tree.  In particular, 
for each subtree problem, $pt_{l,r,b,d}$ stores an ordered tuple $\langle L\_split,branches,depth\_next\rangle$, that captures information about the subtree(s) needed for minimal cost tree construction.

When $b>1$ the tuple effectively decomposes the problem into two sub-problems. $L\_split$ indicates the range of leaves of the first sub-problem as $l$ to $L\_split$ and $branches$ now indicates how many of the branches are in the first sub-problem.  The second sub-problem takes the remaining leaves and branches. The solutions to these two sub-problems are stored in other $pt$ points, namely $pt[l][L\_split][branches][depth\_next]$ and $pt[L\_split+1][r][b-branches][depth\_next]$. 

As an example, for Fig. \ref{fig:pt} the stored result would be $\langle C,1,4\rangle$ indicating that in the solution for the splitter subtree under evaluation (indicated at the dashed line), the first sub-problem contains only node $C$ and the second sub-problem should consider the optimal construction of connecting $1$ branch from level $4$ to nodes $D$ and $E$. This indicates the solution for the second sub-problem can be found at $pt[D][E][1][4]$.\footnote{Note that in this paper the fanout nodes are labelled with single capital letters for ease of exposition and the order and range of nodes is thus defined lexicographically,}
      
When $b=1$, the examined problem has 1 branch feeding into it and $branches$ indicates how many branches are in the next sub-problem used to construct the splitter tree. If $branches = 1$ a buffer is to be inserted, otherwise a splitter must be inserted to satisfy the increase in fanout. $depth\_next$ indicates the next level of the subtree that should be examined such that the next sub-problem solution is in $pt[l][r][branches][depth\_next]$.  The minimal solution stored in $pt[D][E][1][4]$ would be \{-1,1,6\} indicating that a buffer is to be inserted at level $6$ (Fig. \ref{fig:ptopt}) and to jump to sub-problem $pt[D][E][1][6]$. Buffer and splitter locations can be discovered by traversing \textbf{pt} from the root at $pt[A][Z][1][0]$, where $Z$ is the last fanout node, recovering the minimal cost splitter tree.

The construction of \textbf{pt} can be seen in 
lines \ref{pt1} and \ref{pt2} of Algorithm \ref{splitters}, where our backtracing algorithm differs from \cite{dp_OG} and \cite{SOTA} because phase-skipping allows the minimal subtree(s) to be constructed from subtree(s) within a range of depths.  Key phase-skipping modifications include Lines \ref{cost}, which calculates the inserted buffer costs considering phase-skipping, and \ref{span}, which allows for phase-skipping within the splitter tree by exploring a span of depths from which to construct a minimal cost solution. Line \ref{slack} allows for a given node to exceed its set level according to its given slack at no cost. 

 \setcounter{table}{1} 
\begin{table*}[tb]
  \centering
  \caption{Buffer/Splitter and JJ savings for Phase-Skipping Optimization}
    \begin{tabular}{|ccccc|c|c|c|c|c|c|c|c|}
    \hline
    \multicolumn{1}{|c|}{\multirow{5}[8]{*}{\textbf{Benchmark }}} &
      \multicolumn{2}{c|}{\multirow{4}[6]{*}{\textbf{Original Circuit Netlist}}} &
      \multicolumn{10}{c|}{\textbf{Buffer/Splitter Insertion Method}}
      \bigstrut\\
\cline{4-13}    \multicolumn{1}{|c|}{} &
      \multicolumn{2}{c|}{} &
      \multicolumn{2}{c|}{\multirow{2}[2]{*}{\textbf{SOTA \cite{SOTA} }}} &
      \multicolumn{8}{c|}{\multirow{2}[2]{*}{\textbf{Phase-Skipping Optimization Algorithm}}}
      \bigstrut[t]\\
    \multicolumn{1}{|c|}{} &
      \multicolumn{2}{c|}{} &
      \multicolumn{2}{c|}{} &
      \multicolumn{8}{c|}{}
      \bigstrut[b]\\
\cline{4-13}    \multicolumn{1}{|c|}{} &
      \multicolumn{2}{c|}{} &
      \multicolumn{2}{c|}{\textbf{0-Skip}} &
      \multicolumn{2}{c|}{\textbf{1-Skip}} &
      \multicolumn{2}{c|}{\textbf{2-Skip}} &
      \multicolumn{2}{c|}{\textbf{3-Skip}} &
      \multicolumn{2}{c|}{\textbf{4-Skip}}
      \bigstrut\\
\cline{2-13}    \multicolumn{1}{|c|}{} &
      \multicolumn{1}{p{2.445em}}{\textbf{Gates}} &
      \multicolumn{1}{c|}{\textbf{JJs}} &
      \multicolumn{1}{c}{\textbf{BS}} &
      \textbf{JJs} &
      \multicolumn{1}{c}{\textbf{BS}} &
      \textbf{JJs} &
      \multicolumn{1}{c}{\textbf{BS}} &
      \textbf{JJs} &
      \multicolumn{1}{c}{\textbf{BS}} &
      \textbf{JJs} &
      \multicolumn{1}{c}{\textbf{BS}} &
      \textbf{JJs}
      \bigstrut\\
    \hline
    \multicolumn{1}{|c|}{\textbf{c432}} &
      121 &
      \multicolumn{1}{c|}{726} &
      829 &
      2384 &
      \multicolumn{1}{c}{416} &
      1558 &
      \multicolumn{1}{c}{296} &
      1318 &
      \multicolumn{1}{c}{223} &
      1172 &
      \multicolumn{1}{c}{188} &
      1102
      \bigstrut\\
\cline{1-1}    \multicolumn{1}{|c|}{\textbf{c499}} &
      387 &
      \multicolumn{1}{c|}{2322} &
      1173 &
      4668 &
      \multicolumn{1}{c}{665} &
      3652 &
      \multicolumn{1}{c}{491} &
      3304 &
      \multicolumn{1}{c}{401} &
      3124 &
      \multicolumn{1}{c}{356} &
      3034
      \bigstrut\\
\cline{1-1}    \multicolumn{1}{|c|}{\textbf{c880}} &
      306 &
      \multicolumn{1}{c|}{1836} &
      1536 &
      4908 &
      \multicolumn{1}{c}{745} &
      3326 &
      \multicolumn{1}{c}{486} &
      2808 &
      \multicolumn{1}{c}{356} &
      2548 &
      \multicolumn{1}{c}{273} &
      2382
      \bigstrut\\
\cline{1-1}    \multicolumn{1}{|c|}{\textbf{c1355}} &
      389 &
      \multicolumn{1}{c|}{2334} &
      1186 &
      4706 &
      \multicolumn{1}{c}{666} &
      3666 &
      \multicolumn{1}{c}{485} &
      3304 &
      \multicolumn{1}{c}{403} &
      3140 &
      \multicolumn{1}{c}{358} &
      3050
      \bigstrut\\
\cline{1-1}    \multicolumn{1}{|c|}{\textbf{c1908}} &
      289 &
      \multicolumn{1}{c|}{1734} &
      1253 &
      4240 &
      \multicolumn{1}{c}{651} &
      3036 &
      \multicolumn{1}{c}{442} &
      2618 &
      \multicolumn{1}{c}{359} &
      2452 &
      \multicolumn{1}{c}{299} &
      2332
      \bigstrut\\
\cline{1-1}    \multicolumn{1}{|c|}{\textbf{c2670}} &
      369 &
      \multicolumn{1}{c|}{2216} &
      1869 &
      5954 &
      \multicolumn{1}{c}{892} &
      4000 &
      \multicolumn{1}{c}{597} &
      3410 &
      \multicolumn{1}{c}{441} &
      3098 &
      \multicolumn{1}{c}{341} &
      2898
      \bigstrut\\
\cline{1-1}    \multicolumn{1}{|c|}{\textbf{c3540}} &
      794 &
      \multicolumn{1}{c|}{4764} &
      1963 &
      8690 &
      \multicolumn{1}{c}{980} &
      6724 &
      \multicolumn{1}{c}{686} &
      6136 &
      \multicolumn{1}{c}{567} &
      5898 &
      \multicolumn{1}{c}{492} &
      5748
      \bigstrut\\
\cline{1-1}    \multicolumn{1}{|c|}{\textbf{c5315}} &
      1317 &
      \multicolumn{1}{c|}{7932} &
      5505 &
      18942 &
      \multicolumn{1}{c}{2717} &
      13366 &
      \multicolumn{1}{c}{1825} &
      11582 &
      \multicolumn{1}{c}{1383} &
      10698 &
      \multicolumn{1}{c}{1160} &
      10252
      \bigstrut\\
\cline{1-1}    \multicolumn{1}{|c|}{\textbf{c6288}} &
      1870 &
      \multicolumn{1}{c|}{11220} &
      8832 &
      28884 &
      \multicolumn{1}{c}{4681} &
      20582 &
      \multicolumn{1}{c}{3473} &
      18166 &
      \multicolumn{1}{c}{3005} &
      17230 &
      \multicolumn{1}{c}{2477} &
      16174
      \bigstrut\\
\cline{1-1}    \multicolumn{1}{|c|}{\textbf{c7552}} &
      1395 &
      \multicolumn{1}{c|}{8372} &
      6768 &
      21908 &
      \multicolumn{1}{c}{3323} &
      15018 &
      \multicolumn{1}{c}{2262} &
      12896 &
      \multicolumn{1}{c}{1776} &
      11924 &
      \multicolumn{1}{c}{1447} &
      11266
      \bigstrut\\
\cline{1-1}    \multicolumn{1}{|c|}{\textbf{mult8}} &
      439 &
      \multicolumn{1}{c|}{2634} &
      1681 &
      5996 &
      \multicolumn{1}{c}{866} &
      4366 &
      \multicolumn{1}{c}{628} &
      3890 &
      \multicolumn{1}{c}{539} &
      3712 &
      \multicolumn{1}{c}{445} &
      3524
      \bigstrut\\
\cline{1-1}    \multicolumn{1}{|c|}{\textbf{counter16}} &
      29 &
      \multicolumn{1}{c|}{174} &
      66 &
      306 &
      \multicolumn{1}{c}{34} &
      242 &
      \multicolumn{1}{c}{24} &
      222 &
      \multicolumn{1}{c}{22} &
      218 &
      \multicolumn{1}{c}{19} &
      212
      \bigstrut\\
\cline{1-1}    \multicolumn{1}{|c|}{\textbf{counter32}} &
      82 &
      \multicolumn{1}{c|}{492} &
      156 &
      804 &
      \multicolumn{1}{c}{87} &
      666 &
      \multicolumn{1}{c}{61} &
      614 &
      \multicolumn{1}{c}{58} &
      608 &
      \multicolumn{1}{c}{50} &
      592
      \bigstrut\\
\cline{1-1}    \multicolumn{1}{|c|}{\textbf{counter64}} &
      195 &
      \multicolumn{1}{c|}{1170} &
      351 &
      1872 &
      \multicolumn{1}{c}{200} &
      1570 &
      \multicolumn{1}{c}{143} &
      1456 &
      \multicolumn{1}{c}{135} &
      1440 &
      \multicolumn{1}{c}{118} &
      1406
      \bigstrut\\
\cline{1-1}    \multicolumn{1}{|c|}{\textbf{counter128}} &
      428 &
      \multicolumn{1}{c|}{2568} &
      755 &
      4078 &
      \multicolumn{1}{c}{426} &
      3420 &
      \multicolumn{1}{c}{308} &
      3184 &
      \multicolumn{1}{c}{292} &
      3152 &
      \multicolumn{1}{c}{259} &
      3086
      \bigstrut\\
\cline{1-1}    \multicolumn{1}{|c|}{\textbf{alu32}} &
      1513 &
      \multicolumn{1}{c|}{9078} &
      13976 &
      37030 &
      \multicolumn{1}{c}{6880} &
      22838 &
      \multicolumn{1}{c}{4637} &
      18352 &
      \multicolumn{1}{c}{3532} &
      16142 &
      \multicolumn{1}{c}{2868} &
      14814
      \bigstrut\\
    \hline
    \rowcolor[rgb]{ .906,  .902,  .902} \multicolumn{5}{|c|}{\textbf{Total Savings}} &
      \textbf{47.8\%} &
      \textbf{26.2\%} &
      \textbf{63.3\%} &
      \textbf{34.7\%} &
      \textbf{69.4\%} &
      \textbf{38.2\%} &
      \textbf{74.1\%} &
      \textbf{40.7\%}
      \bigstrut\\
    \hline
    \end{tabular}%
  \label{tab:sota}%
\end{table*}%

\setcounter{table}{0} 
\begin{table}[htbp]
  \centering
  \caption{Phase-Skipping Optimization Savings over Buffer Reduction}
   \begin{adjustbox}{max width=\columnwidth}
    \begin{tabular}{|ccccc|c|c|c|c|}
    \hline
    \multicolumn{1}{|c|}{\multirow{5}[6]{*}{\textbf{Benchmark }}} &
      \multicolumn{8}{c|}{\textbf{Buffer/Splitters Inserted By Method}}
      \bigstrut\\
\cline{2-9}    \multicolumn{1}{|c|}{} &
      \multicolumn{4}{c|}{\multirow{3}[2]{*}{\textbf{Buffer Reduction}}} &
      \multicolumn{4}{c|}{\multirow{3}[2]{*}{\textbf{Our Algorithm}}}
      \bigstrut[t]\\
    \multicolumn{1}{|c|}{} &
      \multicolumn{4}{c|}{} &
      \multicolumn{4}{c|}{}
      \\
    \multicolumn{1}{|c|}{} &
      \multicolumn{4}{c|}{} &
      \multicolumn{4}{c|}{}
      \bigstrut[b]\\
\cline{2-9}    \multicolumn{1}{|c|}{} &
      \multicolumn{1}{c|}{\textbf{1-Skip}} &
      \multicolumn{1}{c|}{\textbf{2-Skip}} &
      \multicolumn{1}{c|}{\textbf{3-Skip}} &
      \textbf{4-Skip} &
      \textbf{1-Skip} &
      \textbf{2-Skip} &
      \textbf{3-Skip} &
      \textbf{4-Skip}
      \bigstrut\\
    \hline
    \multicolumn{1}{|c|}{\textbf{c432}} &
      \multicolumn{1}{c|}{419} &
      \multicolumn{1}{c|}{321} &
      \multicolumn{1}{c|}{276} &
      239 &
      416 &
      296 &
      223 &
      188
      \bigstrut\\
    \hline
    \multicolumn{1}{|c|}{\textbf{c499}} &
      \multicolumn{1}{c|}{702} &
      \multicolumn{1}{c|}{572} &
      \multicolumn{1}{c|}{510} &
      472 &
      665 &
      491 &
      401 &
      356
      \bigstrut\\
    \hline
    \multicolumn{1}{|c|}{\textbf{c880}} &
      \multicolumn{1}{c|}{798} &
      \multicolumn{1}{c|}{627} &
      \multicolumn{1}{c|}{544} &
      304 &
      745 &
      486 &
      356 &
      273
      \bigstrut\\
    \hline
    \multicolumn{1}{|c|}{\textbf{c1355}} &
      \multicolumn{1}{c|}{706} &
      \multicolumn{1}{c|}{580} &
      \multicolumn{1}{c|}{512} &
      477 &
      666 &
      485 &
      403 &
      358
      \bigstrut\\
    \hline
    \multicolumn{1}{|c|}{\textbf{c1908}} &
      \multicolumn{1}{c|}{686} &
      \multicolumn{1}{c|}{542} &
      \multicolumn{1}{c|}{476} &
      425 &
      651 &
      442 &
      359 &
      299
      \bigstrut\\
    \hline
    \multicolumn{1}{|c|}{\textbf{c2670}} &
      \multicolumn{1}{c|}{946} &
      \multicolumn{1}{c|}{740} &
      \multicolumn{1}{c|}{622} &
      568 &
      892 &
      597 &
      441 &
      341
      \bigstrut\\
    \hline
    \multicolumn{1}{|c|}{\textbf{c3540}} &
      \multicolumn{1}{c|}{1137} &
      \multicolumn{1}{c|}{969} &
      \multicolumn{1}{c|}{887} &
      826 &
      980 &
      686 &
      567 &
      492
      \bigstrut\\
    \hline
    \multicolumn{1}{|c|}{\textbf{c5315}} &
      \multicolumn{1}{c|}{2970} &
      \multicolumn{1}{c|}{2346} &
      \multicolumn{1}{c|}{2025} &
      1853 &
      2717 &
      1825 &
      1383 &
      1160
      \bigstrut\\
    \hline
    \multicolumn{1}{|c|}{\textbf{c6288}} &
      \multicolumn{1}{c|}{4805} &
      \multicolumn{1}{c|}{3726} &
      \multicolumn{1}{c|}{3396} &
      3194 &
      4681 &
      3473 &
      3005 &
      2477
      \bigstrut\\
    \hline
    \multicolumn{1}{|c|}{\textbf{c7552}} &
      \multicolumn{1}{r|}{3568} &
      \multicolumn{1}{r|}{2807} &
      \multicolumn{1}{r|}{2427} &
      \multicolumn{1}{r|}{2195} &
      3323 &
      2662 &
      1776 &
      1447
      \bigstrut\\
    \hline
    \multicolumn{1}{|c|}{\textbf{mult8}} &
      \multicolumn{1}{c|}{941} &
      \multicolumn{1}{c|}{765} &
      \multicolumn{1}{c|}{695} &
      662 &
      866 &
      628 &
      539 &
      445
      \bigstrut\\
    \hline
    \multicolumn{1}{|c|}{\textbf{counter16}} &
      \multicolumn{1}{c|}{38} &
      \multicolumn{1}{c|}{35} &
      \multicolumn{1}{c|}{34} &
      34 &
      34 &
      24 &
      22 &
      19
      \bigstrut\\
    \hline
    \multicolumn{1}{|c|}{\textbf{counter32}} &
      \multicolumn{1}{c|}{95} &
      \multicolumn{1}{c|}{90} &
      \multicolumn{1}{c|}{88} &
      88 &
      87 &
      61 &
      58 &
      50
      \bigstrut\\
    \hline
    \multicolumn{1}{|c|}{\textbf{counter64}} &
      \multicolumn{1}{c|}{215} &
      \multicolumn{1}{c|}{202} &
      \multicolumn{1}{c|}{198} &
      198 &
      200 &
      143 &
      135 &
      118
      \bigstrut\\
    \hline
    \multicolumn{1}{|c|}{\textbf{counter128}} &
      \multicolumn{1}{c|}{466} &
      \multicolumn{1}{c|}{439} &
      \multicolumn{1}{c|}{428} &
      428 &
      426 &
      308 &
      298 &
      259
      \bigstrut\\
    \hline
    \multicolumn{1}{|c|}{\textbf{alu32}} &
      \multicolumn{1}{c|}{7128} &
      \multicolumn{1}{c|}{5151} &
      \multicolumn{1}{c|}{4200} &
      3606 &
      6880 &
      4637 &
      3532 &
      2868
      \bigstrut\\
    \hline
    \rowcolor[rgb]{ .906,  .902,  .902} \multicolumn{5}{|c|}{\textbf{Average Buffer/Splitter Savings}} &
      \textbf{6.7\%} &
      \textbf{19.5\%} &
      \textbf{26.6\%} &
      \textbf{31.6\%}
      \bigstrut\\
    \hline
    \end{tabular}%
    \end{adjustbox}
  \label{tab:bcr}%
\end{table}%

In total the time complexity of Algorithm \ref{splitters} is $O(Dn^3\lceil log_X n \rceil XP_s)$.  Where $D$ the level difference between the source and latest fanout node, $X$ is a constant set to the maximum fanout of a splitter and $P_s$ is set by the phase overlap of the applied clocking methodology, typically 1 to 5. 

\section{Experimental Results}\label{experiments}

All circuit comparisons are made using common netlists acquired from \cite{benchmarks}.  Currently, the only other way to utilize phase-skipping is buffer chain reduction proposed in \cite{Nphase} in which an existing netlist is optimized without phase-skipping and any inserted buffers that would be redundant under phase-skipping are removed.  
We apply this method to our benchmark circuits and compare the total buffer and splitter costs to our iterative algorithm for 1, 2, 3, and 4 phase-skipping enabled circuits in Table \ref{tab:bcr}.  
Our results demonstrate the saving benefits of phase-skipping netlist optimization as we achieve 6.7\%, 19.5\%, 26.6\% and 31.6\% average savings over buffer chain reduction for 1, 2, 3, and 4 phase-skipping circuits respectively.  These savings demonstrate the impact of our approach to joint buffer and splitter netlist optimization for phase-skipping circuits.  

In Table \ref{tab:sota}, we demonstrate the impact of our algorithm by comparing against benchmark results for the reported state of the art values from \cite{SOTA}, for both total buffer and splitter count (BS) and total JJ count (JJs).  
By exploiting phase-skipping clocking schemes our algorithm on average saves 47.8\%, 63.3\%, 69.4\%, and 74.1\% of total BS count for circuits with 1, 2, 3, and 4 phase-skips enabled.  These buffer and splitter savings result in average savings of 26.3\%, 34.7\%, 38.2\%, and 40.7\% in total circuit JJ count for 1, 2, 3 and 4 phase-skips respectively.  Critically, for the largest circuit in the benchmark alu32, we save 60\% of the total JJ count by using 4 phase-skipping circuits.      

\section{Conclusions}

Despite the advances in latency associated with phase-skipping AQFP circuits, their area efficiency remains an important problem. To this end we present an algorithm that utilizes phase-skipping in buffer and splitter insertion to achieve an average 74\% buffer and splitter reduction for 4 phase-skipping circuits and a 40\% reduction in total JJ count compared to the state of the art buffer and splitter insertion algorithms.  We hope the strength of these results motivate continued research into phase-skipping implementations and that the presented algorithm assists in the practical implementation of such circuits. Further optimizations can explore applying phase-skipping B/S insertion for sequential circuits, specifically targeting multi-threading. Additionally, optimizations during physical design can be tailored for phase-skipping connections.

\bibliographystyle{IEEEtran}
\bibliography{bibliography}
\end{document}